\documentclass[showpacs,showkeys,superscriptaddress]{revtex4}%
\usepackage[center,small,it]{subfigure}
\usepackage{doublespace,amsmath,graphicx}
\newcommand{\f}{$\times 10^{16}$}
\begin{document}
\begin{spacing}{1}
\title{Hierarchical Radiative Quark Mass Matrices with an $U(1)_X$
Horizontal
Symmetry Model\footnote{Published:Rev. Mex. Fis. {\bf 48-1}, 32 (2002)}}

\author{E. Garc\'{\i}a}
\affiliation{Departamento de F\'{\i}sica, Cinvestav \\
Apartado Postal 14-740, 07000 \\
M\'exico, D.F., M\'exico.}
\author{A. Hern\'andez-Galeana}
\affiliation{Departamento de F\'{\i}sica, Escuela Superior de F\'{\i}sica y 
Matem\'aticas \\
Instituto Polit\'ecnico Nacional, U. P. Adolfo L\'opez Mateos \\
M\'exico D.F., 07738, M\'exico }
\author{D. Jaramillo}
\affiliation{Departamento de F\'{\i}sica, Universidad de Antioquia \\
A.A. 1226, Medell\'{\i}n, Colombia.}
\author{W. A. Ponce}
\affiliation{Departamento de F\'{\i}sica, Universidad de Antioquia \\
A.A. 1226, Medell\'{\i}n, Colombia.}
\author{A. Zepeda}
\affiliation{Departamento de F\'{\i}sica, Cinvestav \\
Apartado Postal 14-740, 07000 \\
M\'exico, D.F., M\'exico.}



\begin{abstract}
In a model with a gauge group $G_{SM}\otimes U(1)_X$, where
$G_{SM} \equiv SU(3)_C \otimes SU(2)_L \otimes U(1)_Y$ is the standard
model gauge group and $U(1)_X$ is a horizontal local gauge symmetry, we
propose a radiative generation of the spectrum of quark masses and mixing
angles. The assignment of horizontal charges is such that at tree level only
the third family is massive. Using these tree level masses and
introducing exotic scalars, the light families of quarks
acquire hierarchical masses through radiative corrections. The  rank three 
quark mass matrices obtained are written in terms of a
minimal set of free parameters of the model, whose values are estimated
performing a numerical fit. The resulting quark masses and CKM mixing
angles turn out to be in good agreement with the experimental values.
\end{abstract}

\maketitle

\section{INTRODUCTION} 

Although for several years a great effort has been done to shed some light on the mystery of the fermion masses it still is one of the outstanding puzzles of  particle physics. There have been different approaches to explain the mass hierarchy, the fermion mixing and their possible relation to new physics. A good review covering widely these topics has been presented in \cite{fritzsch}. In this work we will restrict our study to that of a new horizontal symmetry and the derived radiative corrections.

A possible answer to why the masses of the light quarks are so small
compared with the electroweak scale is  that they arise from radiative
corrections\cite{ma}, while the mass of the top quark and possibly those of the
bottom quark and of the tau lepton are generated at tree level. This may be
understood as a consequence of the breaking of a symmetry among families  
(a horizontal symmetry). This symmetry may be discrete\cite{discrete} or
continuous\cite{continuous}. Here we consider the case of a continuous local
horizontal symmetry, a $U(1)_X$ gauge group broken spontaneously.
We limit our calculation of masses to
those of the quark sector, insisting that at tree level only the top and
bottom quarks acquire mass. Instead of assuming a texture for the quark
masses from the beginning, we carry out a one loop and a two
loop calculation of the mass matrices in terms of some parameters which
are the tree level top and bottom quark masses, one Yukawa coupling
and the entries in the mass matrices of the scalar bosons that participate
in the loop diagrams which contribute to
the quark masses.

This paper is organized in the following way: In Section II we
describe explicitly the model, Section III contains the analytical
calculations, while Section IV is devoted to a numerical fit of our equations.
The conclusions are presented in section V.

\section{The Model}
We assume only three families, the Standard Model (SM) families, and we do
not introduce exotic fermions to cancel anomalies. The fermions are
classified as in the SM in five sectors \textit{f} = \textit{q,u,d,l} and
\textit{e}, where \textit{q} and \textit{l} are the $SU(2)_L$ quark and lepton
doublets respectively and \textit{u,d} and \textit{e} are the singlets, in an 
obvious notation. 
In order to reduce the number of parameters and to make the model free
of anomalies, we demand that the values X of the horizontal charge satisfy
the traceless condition\cite{ZEP1}

\begin{equation}\label{Eq. 1}
X(f_i)=0,\pm \delta_f,
\end{equation}
where $i=1,2,3$ is a family index, with the constraint 
\begin{equation}\label{Eq. 2}
\delta_q^2 - 2 \delta_u^2 + \delta_d^2 = \delta_l^2 - \delta_e^2.
\end{equation}
Eq. (\ref{Eq. 1}) guarantees the cancellation of the [U(1)$_H$]$^3$
anomaly as well as those which are linear in the U(1)$_H$ hypercharge
($[SU(3)_c]^2U(1)_H,\; [SU(2)_L]^2U(1)_H,\; [Grav]^2U(1)_H$ and
$[U(1)_Y]^2U(1)_H$). Eq. (\ref{Eq. 2})
is the condition for the cancellation of the U(1)$_Y$[U(1)$_H$]$^2$
anomaly. A solution of Eq. (\ref{Eq. 2}) which guarantees
that only the top and bottom quarks get masses at tree level is given by
(``doublets independent of singlets", see Ref. \cite{ZEP1})  
\begin{equation}\label{Eq. 3}
 \delta_l = \delta_q = \pm\Delta \neq \delta_u = \delta_d = \delta_e
=\pm\delta.
\end{equation}
To  avoid tree level flavor changing neutral currents,  we do not allow
mixing between the standard model Z boson
and its horizontal counterpart. Consequently the SM Higgs scalar should have zero
horizontal charge. As a consequence, and since we insist in having a
non-zero tree-level mass
for the top and bottom quarks, the horizontal charges of these quarks 
should satisfy
\begin{eqnarray}
-X(q_3) + X(u_3) = 0, & -X(q_3) + X(d_3) = 0
\end{eqnarray}
in order for the Yukawa couplings in Eq. (\ref{pyc}) to be invariant,
but then Eqs. (\ref{Eq. 1}) and (\ref{Eq. 3}) demand that they vanish,
\begin{equation}\label{Eq. 4}
 X(u_{3}) = X(q_{3}) = X(d_{3}) = 0,
\end{equation}
which in turn implies $X(l_{3})=X(e_{3})=0$ (this defines the third family). 
The assignment of horizontal charges to the fermions is then as given in
Table 1. The  $SU(3)_C \otimes SU(2)_L \otimes U(1)_Y$ quantum numbers of
the fermions are the same as in the Standard Model.

\begin{table}[!h]
\begin{center}
\begin{tabular}{|l|l|l|l|} \hline
 Sector$\backslash$ Family &    1    &     2    & 3 \\ \hline
 $q$             & $\pm\Delta$ & $\mp\Delta$ & 0 \\
 $u$             & $\pm\delta$ & $\mp\delta$ & 0 \\
 $d$             & $\pm\delta$ & $\mp\delta$ & 0 \\
 $l$             & $\pm\Delta$ & $\mp\Delta$ & 0 \\
 $e$             & $\pm\delta$ & $\mp\delta$ & 0 \\ \hline
\end{tabular}
\end{center}
\caption{\label{tab2} Horizontal charges of fermions.}
\end{table}
\begin{table}[!ht]
\begin{center}
\begin{tabular}{|c|c|c|} \hline
 &\em Class I &\em Class II \\ \hline
   & $\phi_1$\hspace{.3cm} $\phi_2$ & $\phi_3$\hspace{.4cm}
$\phi_4$\hspace{.4cm} $\phi_5$\hspace{.4cm} $\phi_6$\hspace{.4cm}
$\phi_7$\hspace{.4cm} $\phi_8$ \\ \hline 
X & 0 \hspace{.5cm}$-\delta$ & 0
\hspace{.5cm} $\Delta$ \hspace{.4cm} 0 \hspace{.5cm} $\delta$
\hspace{.4cm} 0 \hspace{.5cm} $\delta$ \\ 
Y & 1\hspace{.5cm} 0
&\hspace{-.2cm}$-\frac{2}{3}$\hspace{.3cm} $-\frac{2}{3}$ \hspace{.4cm}
$\frac{4}{3}$\hspace{.5cm} $\frac{4}{3}$\hspace{.3cm}
$-\frac{8}{3}$\hspace{.4cm} $-\frac{8}{3}$ \\ T
&$\frac{1}{2}$\hspace{.5cm} 0 & 1 \hspace{.5cm} 1 \hspace{.5cm} 0
\hspace{.5cm} 0 \hspace{.5cm} 0 \hspace{.5cm} 0 \\ C & 1\hspace{.5cm} 1 &
$\bar{6}$ \hspace{.5cm} $\bar{6}$ \hspace{.5cm} $\bar{6}$ \hspace{.5cm}
$\bar{6}$
\hspace{.5cm} $\bar{6}$
\hspace{.5cm} $\bar{6}$ \\\hline
\end{tabular}
\end{center}
\caption{\label{qnum} Quantum numbers for scalar fields, C denotes the
dimension of the representation under the $SU(3)_c$ color group.}
\end{table}

To generate the first and second family quark masses radiatively we
must introduce new irreducible representations (irreps) of scalar fields,
since the gauge bosons of $G=G_{SM}\otimes U(1)_X$ do not perform
transitions between
different families. Families are of course distinguishable (non
degenerated) only below the scale of
the SM symmetry breaking, when they become massive. 

\vskip.3cm

Looking for scalars which make possible the generation
of fermion masses in a hierarchical manner, we divide the irreps  of
scalar fields into two classes. Class I (II) contains scalar fields
which get (do not get) vacuum expectation value (VEV). 

\vskip.3cm

A proper choice of scalars should be such that no VEVs are induced,
through couplings in the potential, for scalars  in class II. In the model
considered below scalars of class
II have no electrically neutral components, so they never get out of its
class. In our model we introduce two irreps of scalars of class I
and six irreps of scalars of class II, with the quantum numbers
specified in Table \ref{qnum}.
Notice that we introduce just the minimum number of scalars of class I; that is, only
one Higgs doublet of weak isospin to achieve the Spontaneous Symmetry Breaking (SSB) of
the electroweak group down to the
electromagnetic $U(1)_Q$, and one $SU(2)_L$ singlet $\phi_2$ used to break $U(1)_X$. In
this way the horizontal interactions
affect the $\rho$ parameter only at higher orders.

\vskip.3cm

With the above quantum numbers the quark
Yukawa couplings that can be written may be divided into two classes, those of
the D type which are defined by Fig 1a, and
those of the M  type which are defined in Fig 1b. The
Yukawa couplings can thus be written as $L_{Y}
= L_{Y_D} + L_{Y_M}$, where the D Yukawa couplings are

\begin{equation}\label{pyc}
L_{Y_D}  =  Y^u \bar q _{L_3} \tilde{\phi}_1 u_{R_3} + 
Y^d \bar q_{L_3} \phi_1 d _{R_3} + h.c.,
\end{equation}
with $\tilde{\phi}\equiv i\sigma_2\phi^*$, while the M couplings
compatible with the symmetries of the model are
\begin{eqnarray}\label{iyc}
L_{Y_M}=Y_I[q_{1L}^{\alpha T} C \phi_{3 \{ \alpha \beta \} }   
q_{2L}^{\beta} + q_{3L}^{\alpha T} C \phi_{3 \{ \alpha \beta \} }
q_{3L}^{\beta} + q_{2L}^{\alpha T}   
C \phi_{4 \{ \alpha \beta \} } q_{3L}^{\beta} \nonumber\\+ d_{2R}^T C
\phi_5 d_{1R} + d_{3R}^T C \phi_5
d_{3R} + d_{3R}^T C \phi_6 d_{2R} \nonumber\\+ u_{2R}^T C \phi_7 u_{1R} +
u_{3R}^T C
\phi_7 u_{3R} + u_{3R}^T C \phi_8 u_{2R}] + h.c.
\end{eqnarray}
In these couplings C represents the charge conjugation matrix and
$\alpha$ and $\beta$ are weak isospin indices. Color indices
have not been written explicitly. By simplicity and economy we have
assumed only one Yukawa
constant $Y_I$ for all the M couplings. Notice that
$\phi_{3\{\alpha\beta
\}}$
is represented as

\begin{equation}\label{triplet}
\phi_3 = \left( \begin{array}{cc}
\phi^{-4/3} &  \phi^{-1/3} \\
\phi^{-1/3} & \phi^{2/3} \\
\end{array} \right) 
\end{equation}
where the superscript denotes the electric charge of the field.
The same applies for $\phi_4$. 

\vskip.3cm

Scalar fields which are not SU(2)$_L$ doublets do not participate in D type
Yukawa terms, they however contribute to the mass matrix of the scalar
sector and in turn determine the magnitude of the radiatively generated
masses of fermions, as we shall see below.

\vskip.3cm

The most general scalar potential of dimension $\leq 4$ that can be written is 

\begin{eqnarray}\label{Pot4d}
 -V(\phi_i){\mbox{\hspace{5mm}}}= \sum_i \mu_i^2 \mid \phi_i \mid ^2 +
\sum_{i,j}\lambda_{ij}
\mid \phi_i \mid ^2 \mid \phi_j \mid ^2 + 
\eta_{31} \phi_{1}^{\dag} \phi^{\dag}_3 \phi_3 \phi_1 
+\tilde{\eta_{31}} \tilde{\phi_{1}}^{\dag} \phi^{\dag}_3 \phi_3
\tilde{\phi_1} \nonumber \\
 + \eta_{41} \phi_{1}^{\dag} \phi^{\dag}_4 \phi_4 \phi_1 +
\tilde{\eta_{41}} \tilde{\phi_{1}}^{\dag} \phi^{\dag}_4 \phi_4
\tilde{\phi_1}+
 \sum_{ \substack{ i\neq j \\i,j\neq 1,2 }} \eta_{ij}
\mid \phi_i^{\dag} \phi_j \mid^2  + ( \rho_1 \phi_5^{\dag}\phi_6 \phi_2 + 
\nonumber \\
\rho_2 \phi_7^{\dag} \phi_8 \phi_2  + \lambda_1 \phi_5^{\dag}
\phi_1^{\alpha} \phi_{3 \{ \alpha \beta \} }\phi_1^{\beta} + 
 \lambda_2
\phi_7^{\dag}
\tilde{\phi_1}^{\alpha} \phi_{3 \{ \alpha \beta \} }\tilde{%
\phi_1}^{\beta} 
 + \nonumber \\ \lambda_3 Tr(\phi_3^{\dag} \phi_4)
\phi_2^2 + \lambda_4 \phi_5 \phi_6 \phi_7 \phi_2 + 
\lambda_5 \phi_5 \phi_6^{\dag} \phi_7^{\dag} \phi_8 +
h.c. ),
\end{eqnarray}

where $Tr$ means trace and in $ \mid
\phi_i \mid ^2  \equiv \phi_i^{\dag} \phi_i $ an appropriate contraction of
the $SU(2)_L$ and $SU(3)_C$ indices is understood. The gauge invariance of
this potential requires the relation $\Delta = 2\delta$.


\begin{figure}
\begin{center}
\subfigure[]{
\includegraphics[height=2cm]{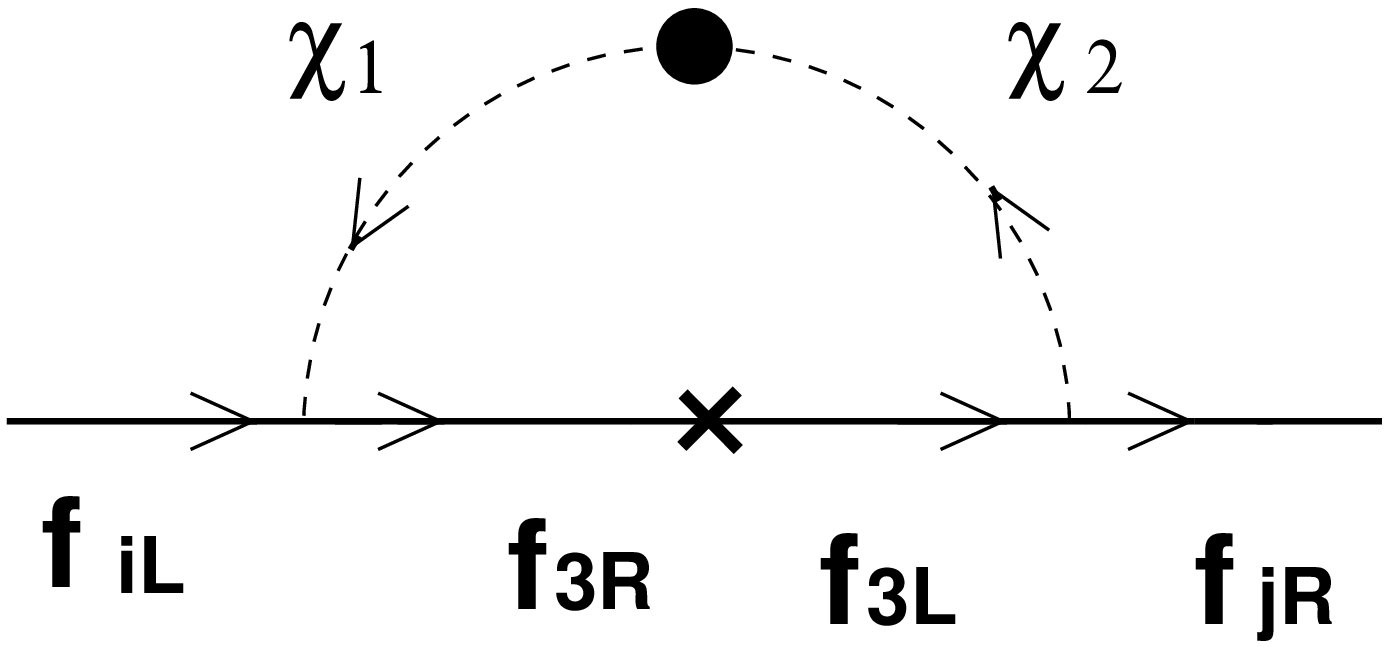}}
\hspace{1.5cm}
\subfigure[]{
\includegraphics[height=2cm]{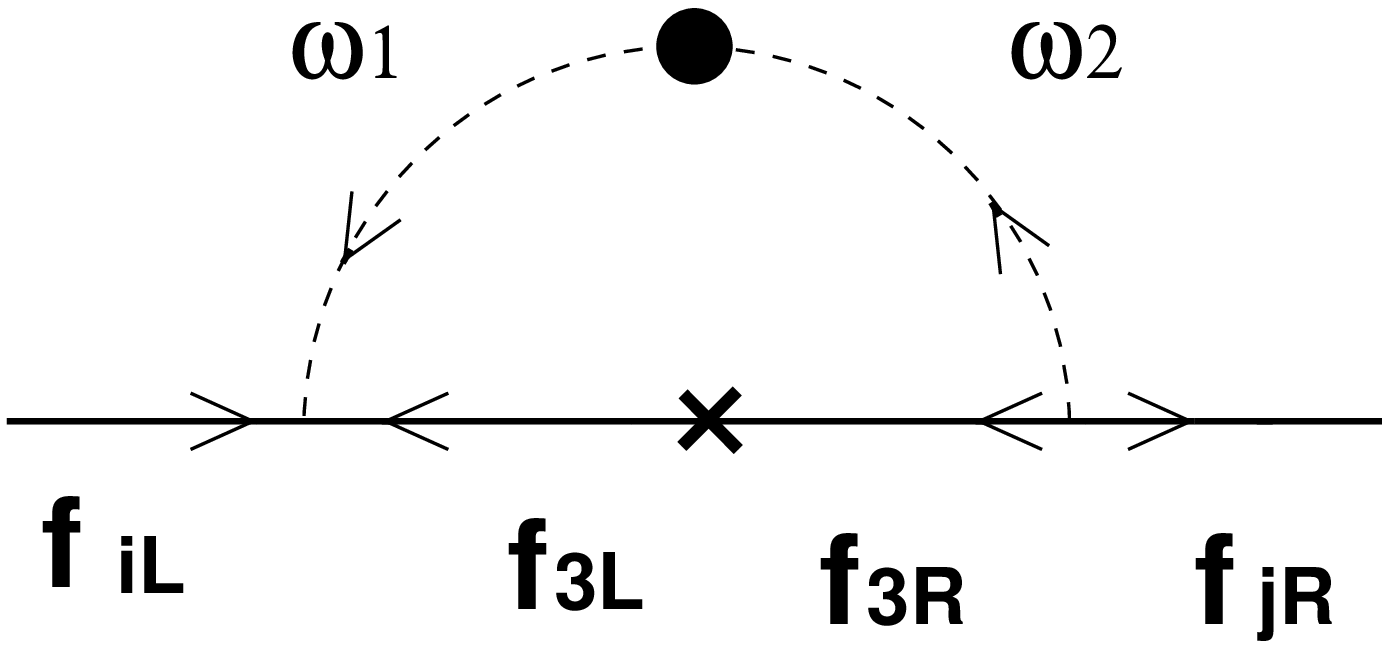}}
\caption{Generic diagrams that could contribute to the mass of the light
families, (a) D type couplings are represented with vertices where
one fermion is incoming and the other one is outgoing, (b) the M    
type couplings are represented with vertices where both fermions are
incoming or outgoing.} \label{M2fig} \end{center}
\end{figure}

\begin{figure}[ht]
\centering
$$
\begin{array}{ccc}
\includegraphics[height=1.8cm]{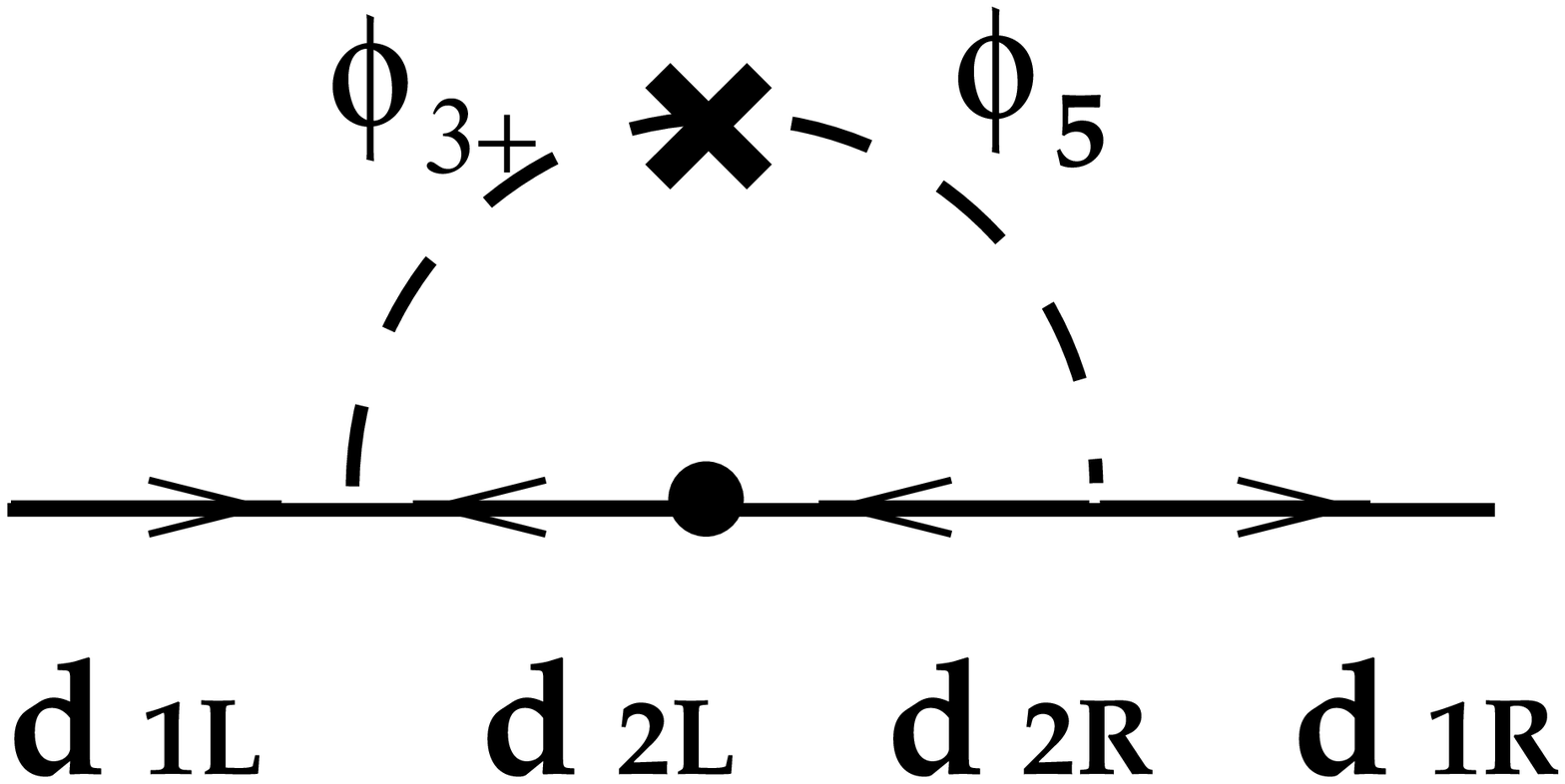} &
\includegraphics[height=1.8cm]{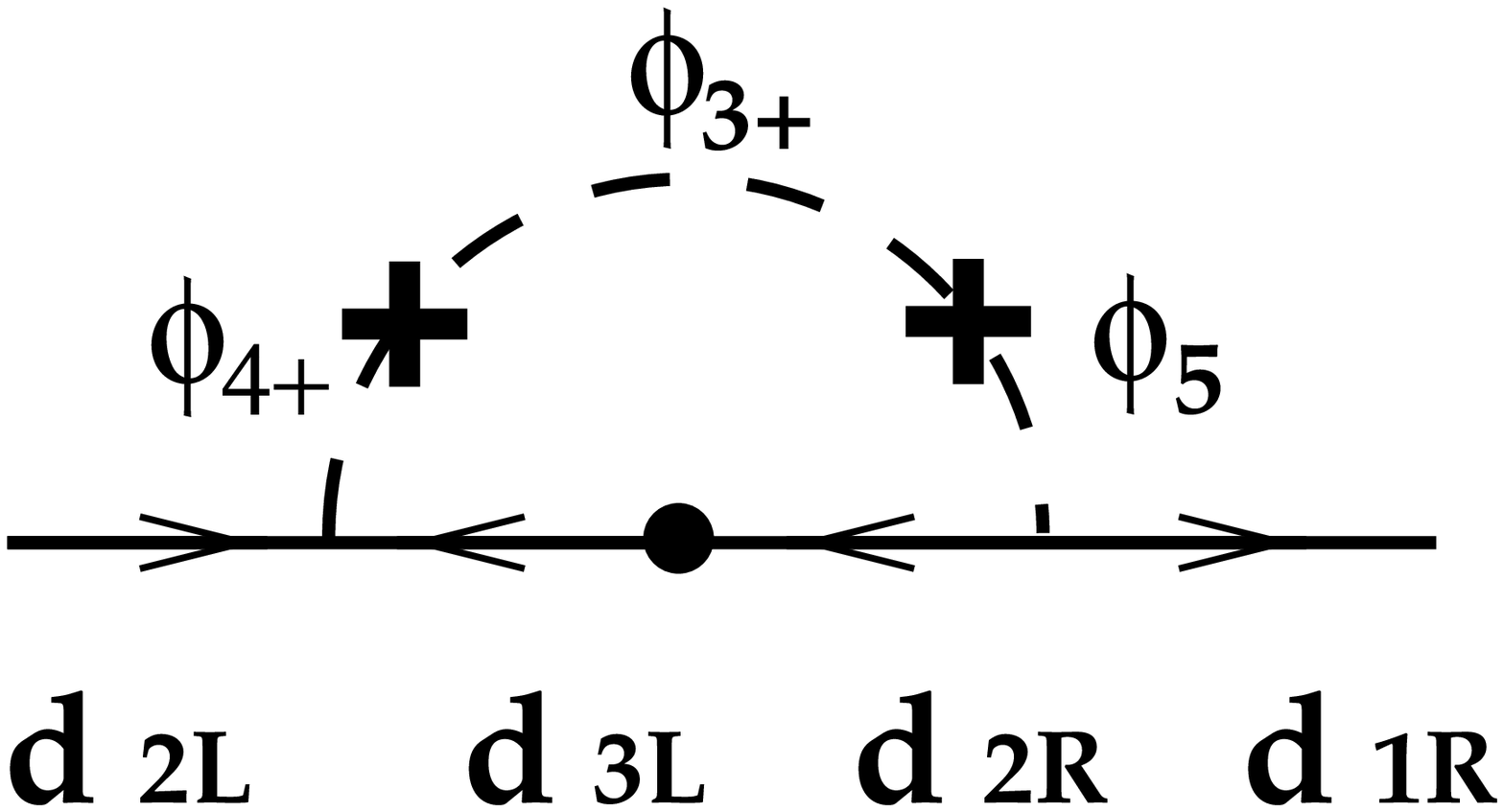} &

\includegraphics[height=1.8cm]{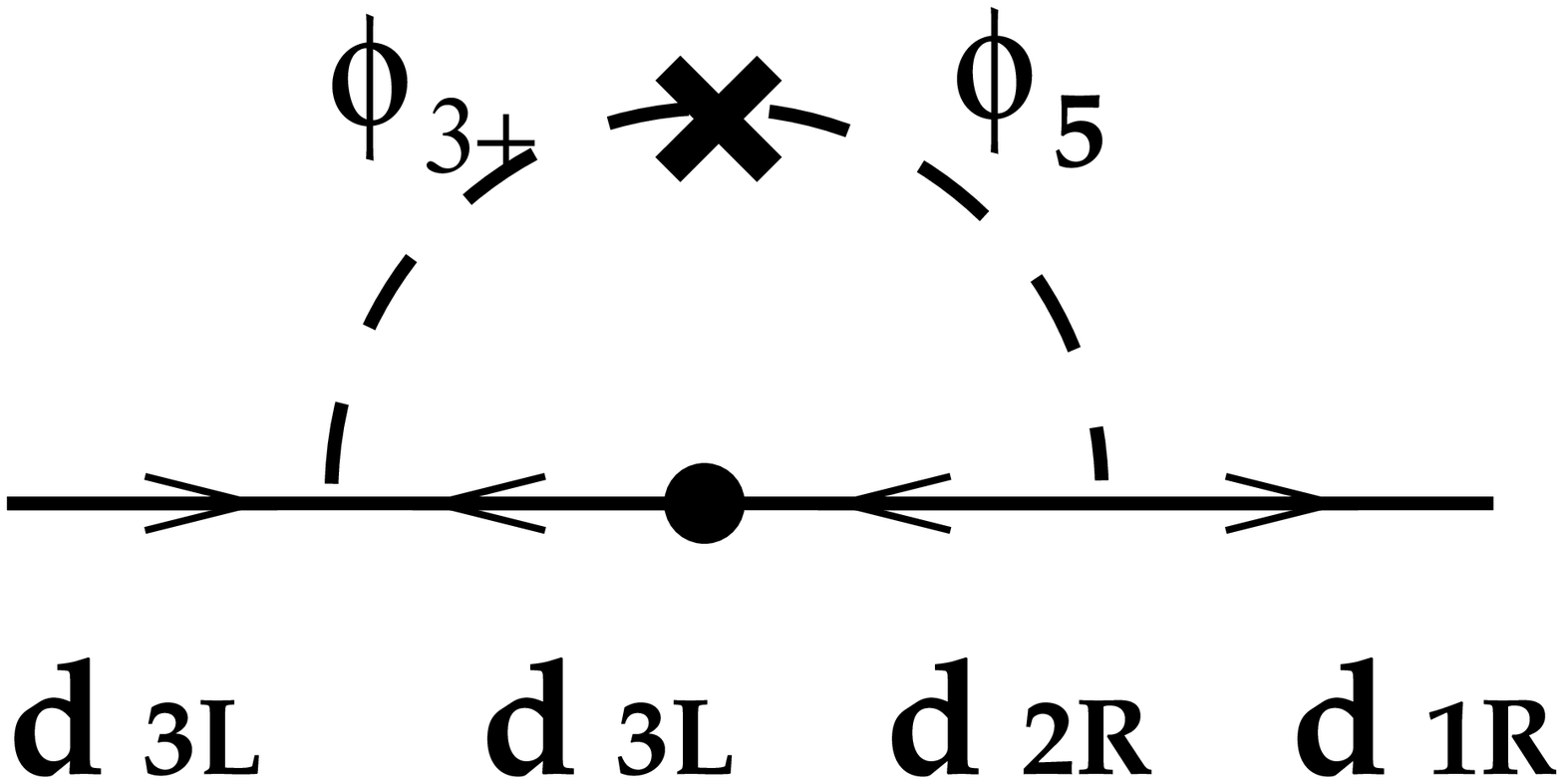}

\\

\includegraphics[height=1.8cm]{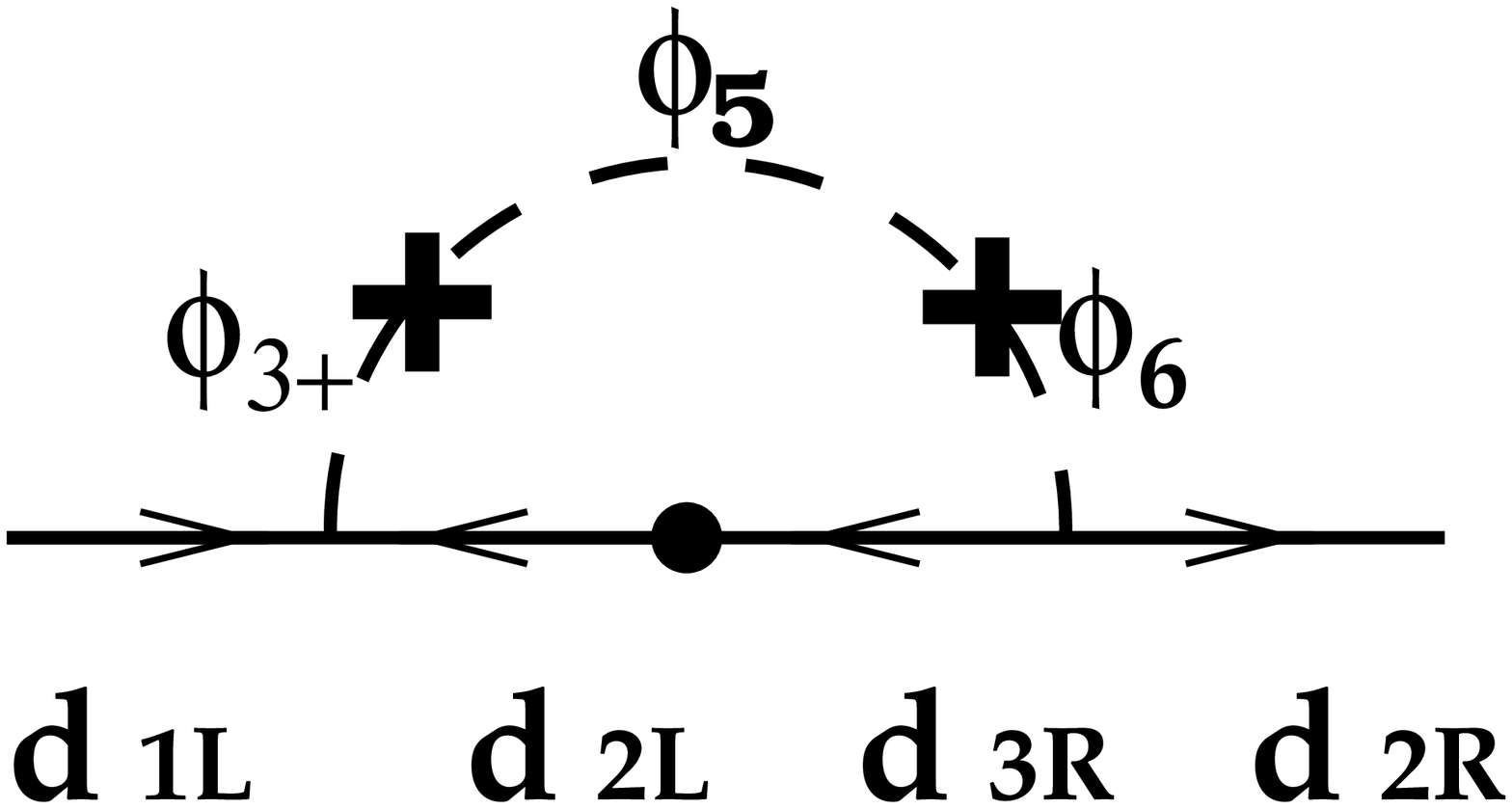} &
\includegraphics[height=1.8cm]{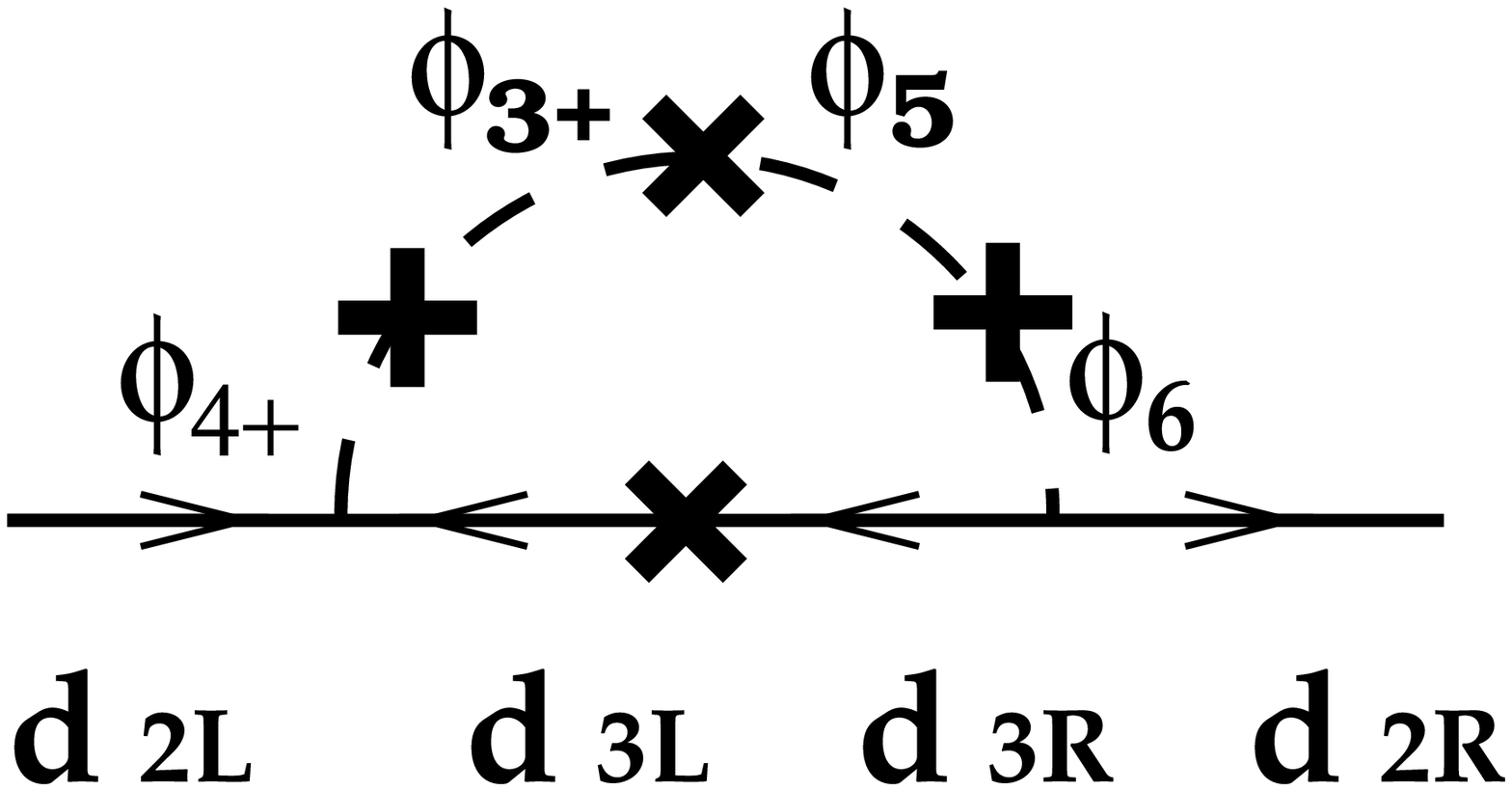} &
\includegraphics[height=1.8cm]{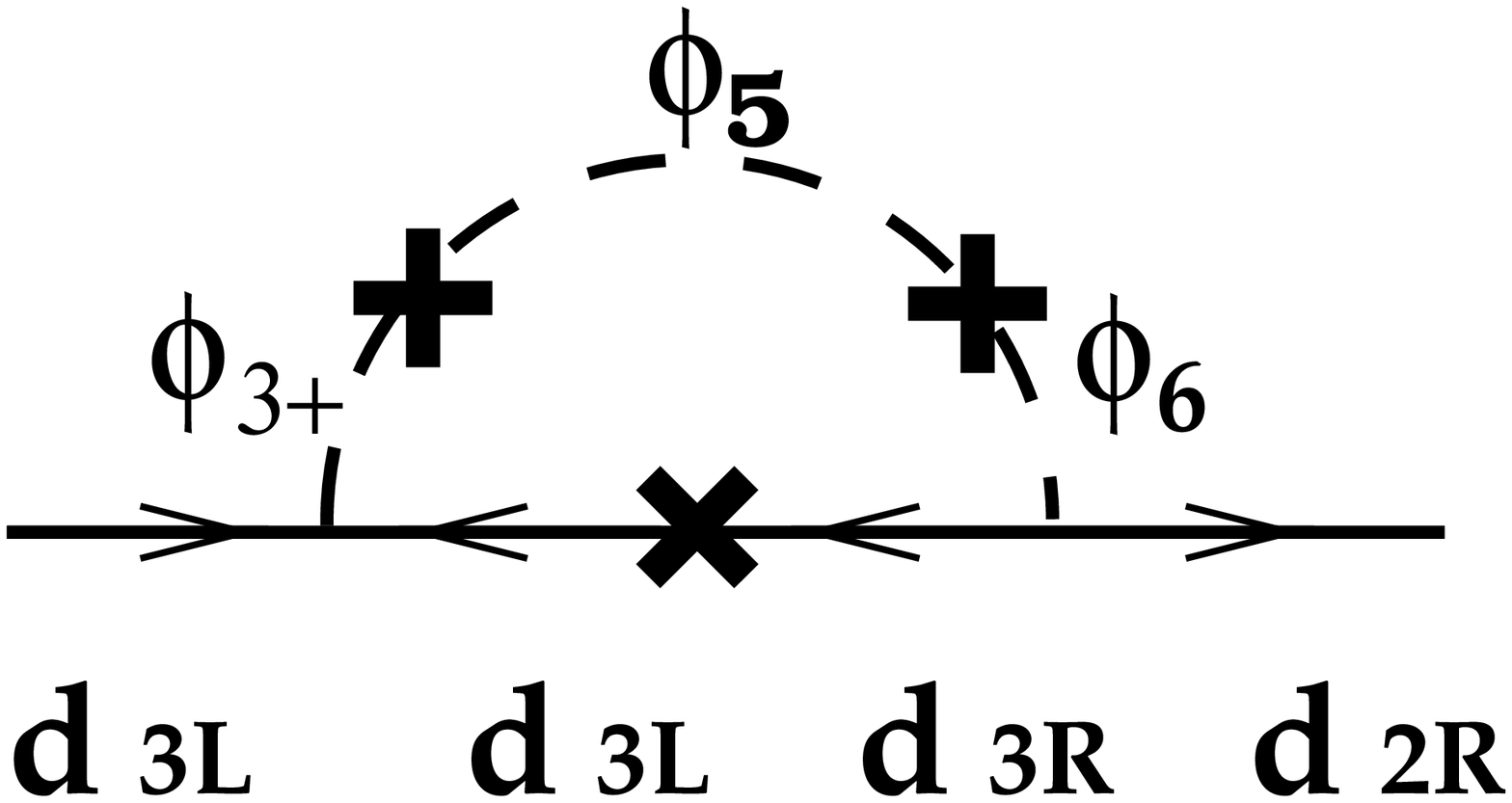} \\

\includegraphics[height=1.8cm]{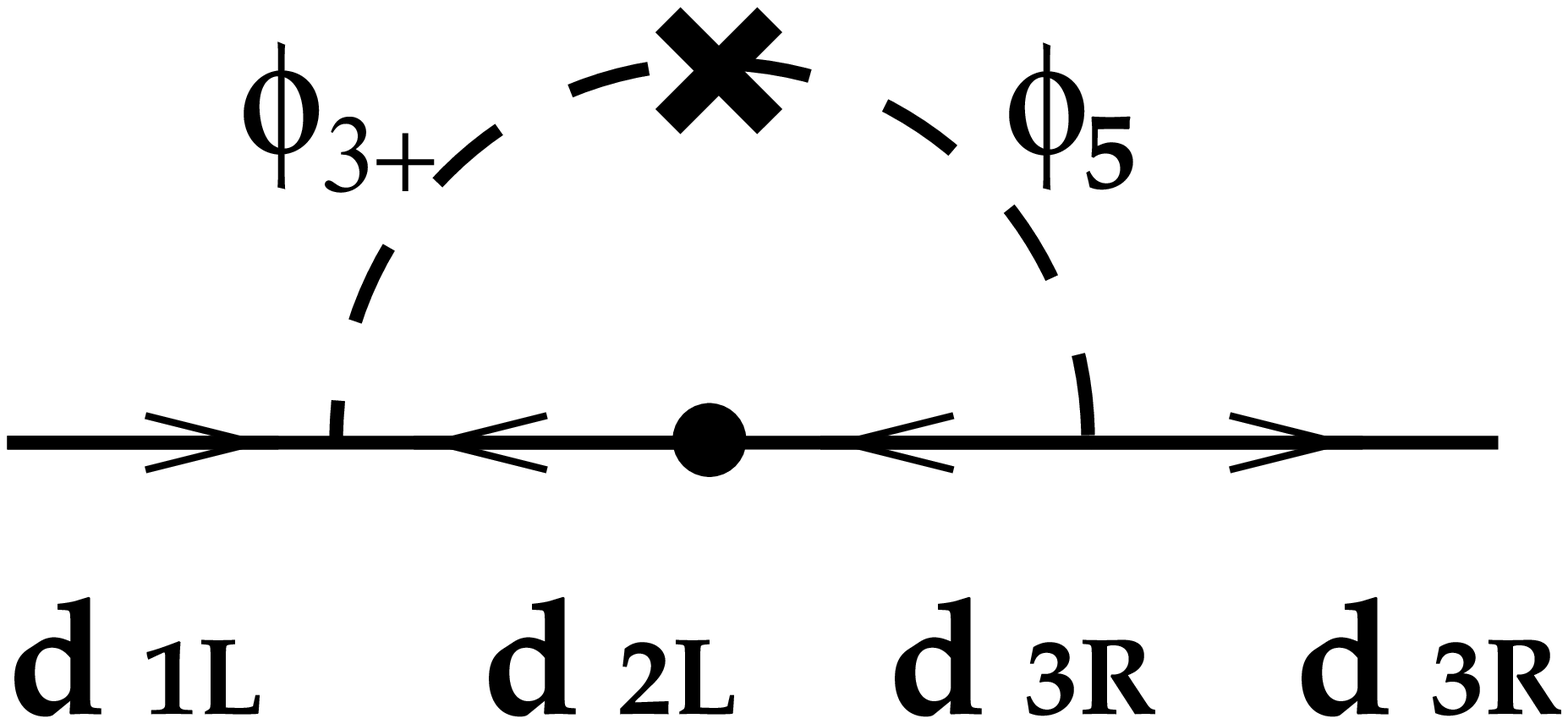}
%
&
\includegraphics[height=1.8cm]{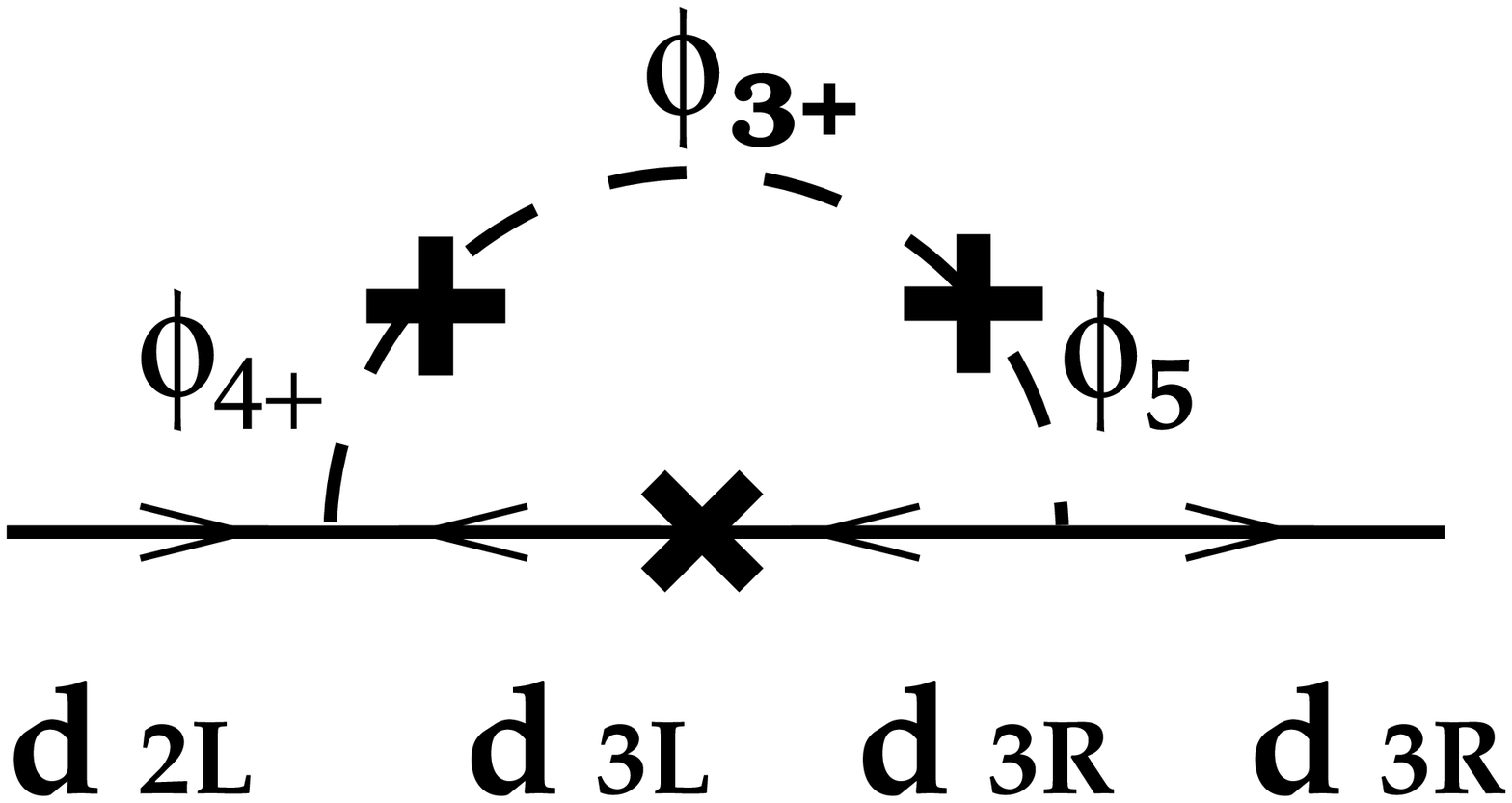} &
\includegraphics[height=0.9cm]{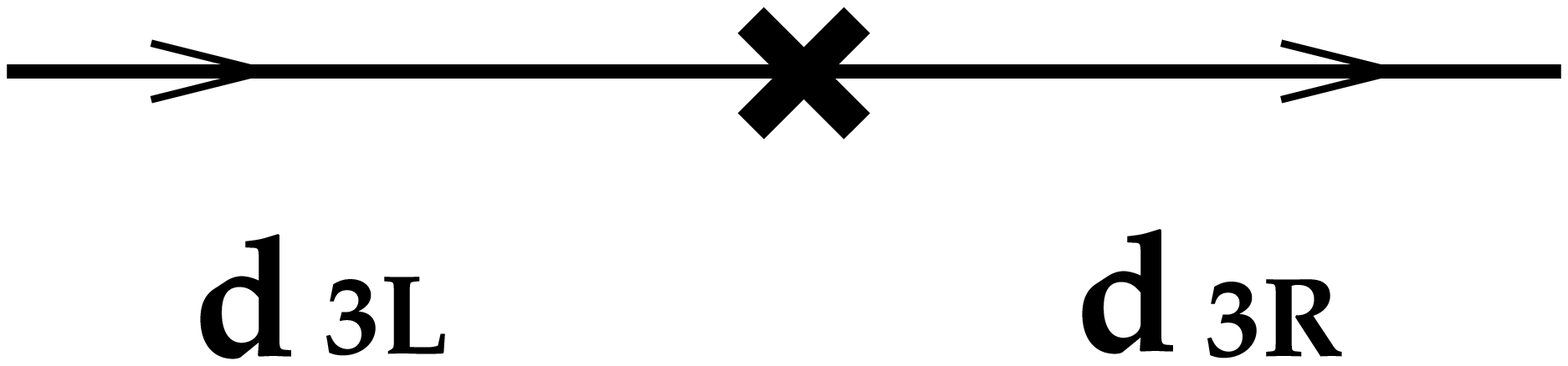}
\end{array} 
$$
\caption{\label{mcd} Mass matrix elements for d quarks}
\end{figure}

\begin{figure}
\begin{center}
\subfigure[]{\centering\includegraphics[height=1.8cm]{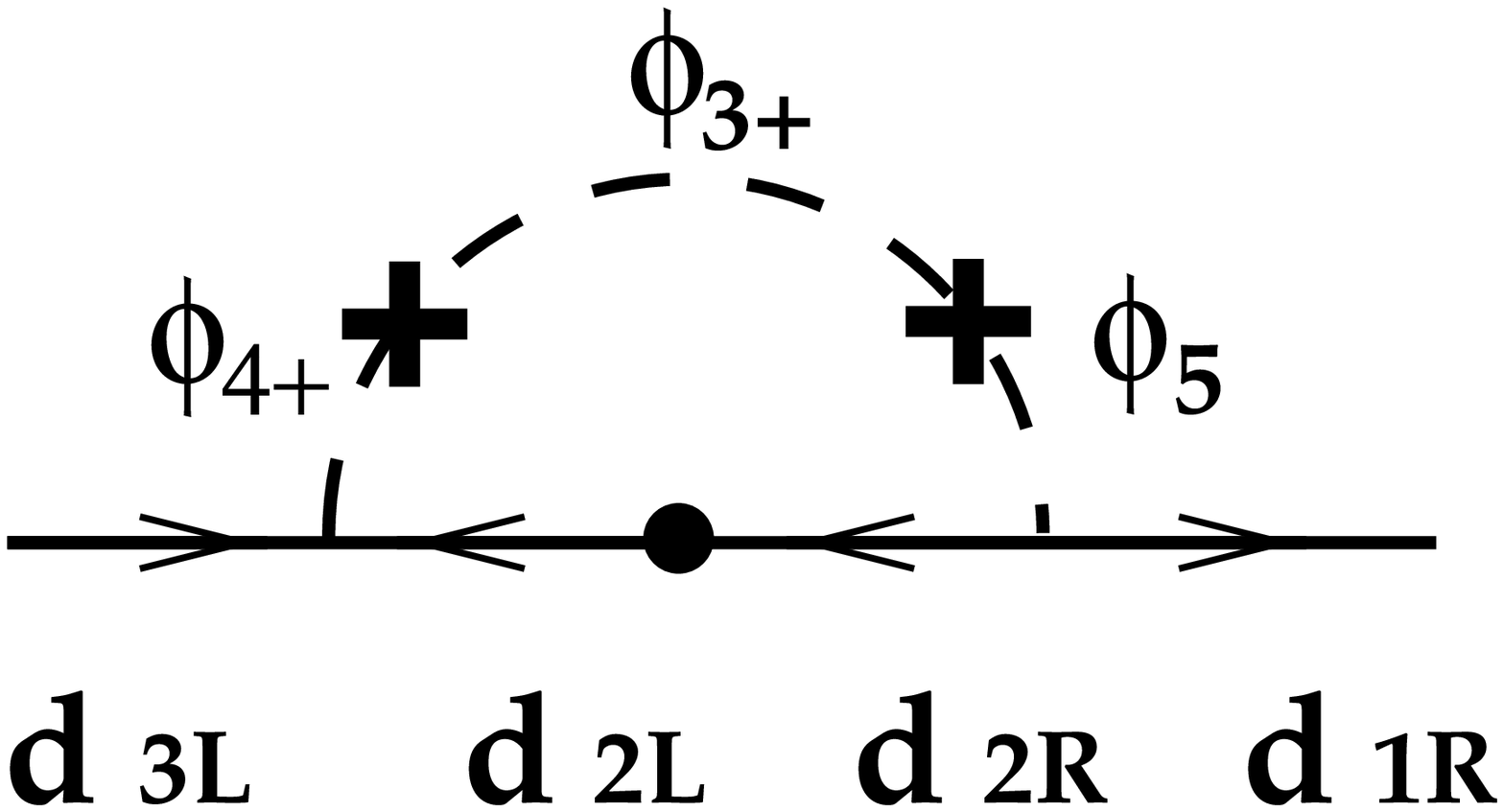}}
\subfigure[]{\centering\includegraphics[height=1.8cm]{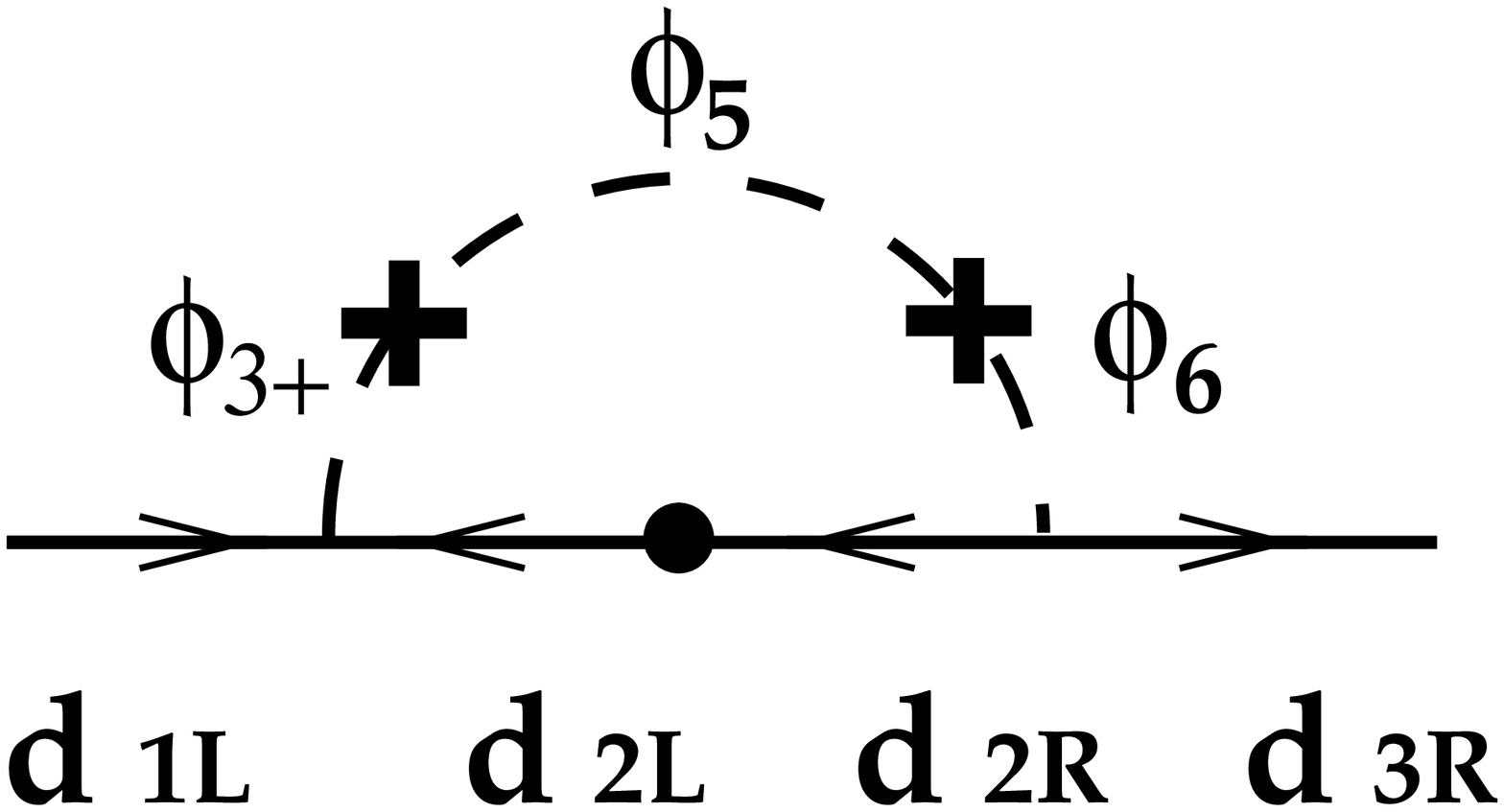}}
\caption{Extra diagrams that contributes to mass matrix elements (a)(1,3)
and (b) (3,1)}
\label{m31b}
\end{center}
\end{figure}

Now we proceed  to describe the mechanism that produces
the quark masses. In general we could have contributions of
two types as depicted in Fig \ref{M2fig}.
In the present model however, we have only the diagrams of Fig
\ref{mcd} for
the charge $-1/3$ quark mass matrix elements and similar ones for the
charge $2/3$ sector (these type of diagrams were first introduced in
\cite{ma}); in those diagrams of Fig \ref{mcd} the cross means
tree
level
mixing and the black circle means one loop mixing. The diagrams in Fig
\ref{m31b}a y \ref{m31b}b should be added to the matrix elements (1,3)
and (3,1), respectively.
In the one loop contribution to the mass matrices for the different quark
sectors only the third family of quarks appears in the internal lines.
This generates a rank 2 matrix, which once diagonalized gives the
physical states at this approximation. Then using these mass eigenstates
we compute the next order contribution, obtaining a matrix of rank 3.
After the diagonalization of this matrix we get the mass eigenvalues and
eigenstates (A quark mass mechanism with some  similar features to
the one proposed here is given in \cite{ma}).

\vskip.3cm

The VEVs of the class I scalar fields are
\begin{eqnarray}
<\phi_1>=\frac{1}{\sqrt{2}}\left( \begin{array}{c} 0 \\ v_1 
\end{array} \right), & <\phi_2>=v_2,
\end{eqnarray}
and they achieve  the breaking
\begin{equation}\label{cadena}
G_{SM}\otimes U(1)_X \stackrel{<\phi_2>}{\longrightarrow} G_{SM}
\stackrel{<\phi_1>}{\longrightarrow} SU(3)_C \otimes
U(1)_Q.
\end{equation}
The scalar field mixing arises after SSB  
from the terms in the potential that couple two different class II fields
to one of class I. After SSB the mass matrices for the scalar fields of
charge $2/3$ ($\phi_{4}$,$\phi_{3}$,$\phi_5$,$\phi_6$) and
$4/3$ ($\phi_{4}$,$\phi_{3}$,$\phi_7$,$\phi_8$) are written,
respectively, as

\begin{eqnarray}\label{scalarmass}
M_{2/3}^2=\left( \begin{array}{cccc}
s_{4}^2   &  \lambda_3^* v_2^2  &      0         &   0 \\
\lambda_3 v_2^2 &  s_{3}^2      & \frac{ \lambda_1^*
v_1^2}{2} & 0 \\
0  & \frac{ \lambda_1 v_1^2}{2} & u_5^2 & \rho_1
v_2 \\
0 & 0 & \rho_1^* v_2 & u_6^2 
\end{array} \right) & \mbox{and} & M_{4/3}^2=\left(
\begin{array}{cccc}
t_{4 }^2   &  \lambda_3^* v_2^2  &      0         &   0 \\ 
\lambda_3 v_2^2 &  t_{3 }^2  & \frac{ \lambda_2
v_1^2}{2} & 0 \\ 
0  & \frac{ \lambda_2^* v_1^2}{2} & t_7^2 & \rho_2
v_2 \\ 
0 & 0 & \rho_2^* v_2 & t_8^2 
\end{array} \right),
\end{eqnarray}
where from Eq. (\ref{Pot4d}) $t_{i}^2 = u_{i}^2 =
\mu_i^2+\lambda_{i1}v_1^2+\lambda_{i2}v_2^2 $  
and $s_{i}^2=t_{i}^2+\eta_{i1}v_1^2$.

Notice that due to the scalar mixing in all the loop diagrams of
Fig \ref{mcd} and \ref{m31b}, the divergences in each one of these
diagrams cancel as is physically expected, giving rise to finite
contributions to the quark mass matrices.

\vskip.3cm

Explicitly, the non vanishing contributions from the diagrams of Fig
\ref{mcd} to the mass terms
$ \bar{d}_{iR} d_{jL} \Sigma^{(1)}_{ij} + h.c.$ read  at
one loop

\begin{equation}\label{sigma22} \Sigma_{22}^{(1)} = 3 m_b^{(0)}
\frac{Y_I^2}{16\pi^2} \sum_k U_{1k}U_{4k}f(M_k,m_b^{(0)}),
\end{equation} 

\begin{equation}\label{sigma23}
\Sigma_{23}^{(1)}=3 m_b^{(0)}\frac{Y_I^2}{16\pi^2}\sum_kU_{2k}U_{4k}
f(M_k,m_b^{(0)}),
\end{equation}

\begin{equation}\label{sigma32}
\Sigma_{32}^{(1)} = 3 m_b^{(0)} \frac{Y_I^2}{16\pi^2} \sum_k
U_{1k}U_{3k}f(M_k,m_b^{(0)}),
\end{equation}
where $m_b^{(0)}$ is the tree level contribution to the b
quark mass, the 3 is a color factor, U is the orthogonal matrix which
diagonalizes the mass matrix
of the charge $2/3$ scalars,

$$(\phi_{4},\phi_{3},\phi_5,\phi_6)^T=U(\sigma_1,\sigma_2,\sigma_3,
\sigma_4)^T,$$ 
where $\sigma_i$ are the eigenfields with eigenvalues $M_i$, and
$$f(a,b) \equiv \frac{1}{a^2-b^2}[a^2ln\frac{a^2}{b^2}],$$
which is just a logarithmic contribution when $a^2 \gg b^2$. The
resulting second rank mass matrix at this level is thus

\begin{equation}\label{mass1}
M_d^{(1)} = \left( \begin{array}{ccc}
 0 & 0 & 0 \\
 0 & \Sigma_{22}^{(1)} & \Sigma_{23}^{(1)} \\
 0 & \Sigma_{32}^{(1)} & m_b^{(0)}
\end{array}  \right).
\end{equation}
At effective two loops we obtain the following expressions:\\
\begin{equation}\label{sigma11.2}
\Sigma_{11}^{(2)} = 3 \frac{Y_I^2}{16\pi^2} \sum_{k,i}m_i^{(1)}
(V^{(1)}_{dL})_{2i}(V^{(1)}_{dR})_{2i}U_{2k}U_{3k}f(M_k,m_i^{(1)}),
\end{equation}  
\begin{equation} \label{sigma12.2}
\Sigma_{12}^{(2)} = 3 \frac{Y_I^2}{16\pi^2} \sum_{k,i}m_i^{(1)}
(V^{(1)}_{dL})_{3i}(V^{(1)}_{dR})_{2i}U_{1k}U_{3k}f(M_k,m_i^{(1)}),
\end{equation}

\begin{eqnarray}\label{sigma13.2}
\Sigma_{13}^{(2)} = 3 \frac{Y_I^2}{16\pi^2} \sum_{k,i}m_i^{(1)}
(V^{(1)}_{dL})_{3i}(V^{(1)}_{dR})_{2i}U_{2k}U_{3k}f(M_k,m_i^{(1)})
\\\nonumber
+
3 \frac{Y_I^2}{16\pi^2} \sum_{k,i}m_i^{(1)}
(V^{(1)}_{dL})_{2i}(V^{(1)}_{dR})_{2i}U_{1k}U_{3k}f(M_k,m_i^{(1)}),
\end{eqnarray} 

\begin{equation}\label{sigma21.2}
\Sigma_{21}^{(2)} =3 \frac{Y_I^2}{16\pi^2} \sum_{k,i}m_i^{(1)}
(V^{(1)}_{dL})_{2i}(V^{(1)}_{dR})_{3i}U_{2k}U_{4k}f(M_k,m_i^{(1)}),
\end{equation} 

\begin{eqnarray}\label{sigma31.2}
\Sigma_{31}^{(2)} = 3 \frac{Y_I^2}{16\pi^2} \sum_{k,i}m_i^{(1)}
(V^{(1)}_{dL})_{2i}(V^{(1)}_{dR})_{3i}U_{2k}U_{3k}f(M_k,m_i^{(1)})
\\\nonumber
+
3 \frac{Y_I^2}{16\pi^2} \sum_{k,i}m_i^{(1)}
(V^{(1)}_{dL})_{2i}(V^{(1)}_{dR})_{2i}U_{2k}U_{4k}f(M_k,m_i^{(1)}),
\end{eqnarray} 
where the k (i) index goes from 1 to 4 (from 2 to 3),
$V^{(1)}_{dL}$ and $V^{(1)}_{dR}$ are the unitary matrices which
diagonalize $M_d^{(1)}$ of equation
(\ref{mass1}) and $m_i^{(1)}$ are the eigenvalues.
Therefore at two loops the mass matrix for d quarks becomes:

\begin{equation}\label{mass2}
M_d^{(2)} = \left( \begin{array}{ccc}
\Sigma_{11}^{(2)} & \Sigma_{12}^{(2)} & \Sigma_{13}^{(2)} \\
\Sigma_{21}^{(2)} & m_2^{(1)} & 0 \\
\Sigma_{31}^{(2)} & 0 &  m_3^{(1)} 
\end{array} \right).
\end{equation}

For the up sector the procedure to obtain the masses is completely
analogous. That is, the mass terms for the up sector come from graphs like
those in Fig. \ref{mcd} and \ref{m31b}, but replacing the
$\phi_{4}$,$\phi_{3}$, $\phi_5$ and $\phi_6$ scalar fields by
$\phi_{4}$,$\phi_{3}$, $\phi_7$ and $\phi_8$ and the quarks $d_i$ by the
quarks $u_i$.

\vskip.3cm

The CKM matrix takes the form

\begin{equation}\label{ckmmatrix}
V_{CKM} = (V_{uL}^{(2)} V_{uL}^{(1)})^{\dag} V_{dL}^{(2)} V_{dL}^{(1)},
\end{equation}
where the unitary matrices $V_{uL}^{(1)}$ and $V_{uR}^{(1)}$ diagonalize
$M_u^{(1)}$, and $V_{uL}^{(2)}$ and $V_{uR}^{(2)}$ diagonalize
$M_u^{(2)}$, with an analogous notation used for the down sector.

It is important to mention here that the textures, particularly the
zeros in the scalar and quark mass matrices
(Eqs. \ref{scalarmass} and \ref{mass2}) are not accidental neither
imposed; they are just a direct consequence of
the mass mechanism that we are introducing and of the gauge symmetry of
the model.

\section{Numerical Evaluation}
\subsection{Experimental values}
Since quarks are confined inside hadrons, their masses can not be directly measured. So, the quark
mass parameters in the SM Lagrangian depend both on the renormalization scheme adopted to define
the theory and on the scale parameter $\mu$ where the theory is being tested. In the limit where
all quark masses are zero, the SM has an $SU(3)_L\otimes SU(3)_R$ chiral symmetry which is broken
at an scale $\Lambda_\chi\simeq 1$GeV. To determine the quark mass values one must use 
SM perturbation theory at an energy scale $\mu>>\Lambda_\chi$ where non perturbative
effects are negligible.

For illustration, the allowed ranges of quark masses\cite{pdg} in the modified 
minimal subtraction scheme $(\overline{MS})$ are\cite{nari}:
\begin{eqnarray}\nonumber
m_u(1.GeV)&=&2-6.8\; MeV.\\ \nonumber 
m_d(1.GeV)&=&4-12\; MeV.\\ \nonumber
m_s(1.GeV)&=&81-230\; MeV.\\  \nonumber
m_c(m_c)&=&1.1-1.4\; GeV.\\ \nonumber
m_b(m_b)&=&4.1-4.4\; GeV.\\ \nonumber
m_t(Exp)&=&173.8\pm 5.2 GeV 
\end{eqnarray}
To get the relative magnitude of different quark masses in a meaningful way, one has
to describe all quark masses in the same scheme and at the same scale. In our analysis
we are calculating the quark masses at an energy scale $\mu_m$ such
that $M_Z<\mu_m<M_X\simeq v_2$, where
$M_X$ is the mass scale where $U(1)_X$ is spontaneously broken. Since in our model
there is no mixing between the Standard Model Z boson and its horizontal counterpart,
we can have $v_2$ as low as the electroweak breaking scale.
For simplicity, let us assume that our calculations are meaningful at the
electroweak breaking
scale and from the former values for the quark masses let us calculate, in the
$\overline{MS}$
scheme, the quark masses at the $m_t$ scale\cite{fusaoka} and at the $M_Z$
scale\cite{fritzsch}.
Those values calculated in the references cited, are presented in tables \ref{ten:res}  
and \ref{nein:res} respectively.

On the other hand, the CKM matrix elements are not ill defined and they can be directly measured
from the charged weak current in the SM. For simplicity we assume that they are real, and as
discussed in Ref.\cite{fusaoka}, they are almost constant in the interval $M_Z<\mu<$ a few TeV.
Their current experimental value\cite{pdg} are given in the Tables \ref{ten:res} and
\ref{nein:res}.

\subsection{Evaluation of the parameters}
In order to test the model using the least possible number of free
parameters, let us write the scalar mass matrices in the following form: 

\begin{eqnarray}\label{emm}
M_{2/3}^2 = \left(
\begin{array}{cccc}
a_+ & b & 0        & 0 \\
b & a_+ & c_+  & 0 \\
0 & c_+ & a_+  & d_+ \\
0 & 0 & d_+ & a_+
\end{array} \right)
& \mbox{and}\hspace{1cm} &
M_{4/3}^2 = \left(
\begin{array}{cccc}
a_- & b & 0        & 0 \\
b & a_- & c_-  & 0 \\
0 & c_- & a_-  & d_- \\ 
0 & 0 & d_- & a_-
\end{array} \right).
\end{eqnarray} \\
Using the central value of the CKM elements in the PDG book\cite{pdg} and the central values
of the six quark masses at the top mass scale\cite{fusaoka},
we build the $\chi^2$ function in the ten parameter space defined by
$(a_+,a_-,b,c_+,c_-,d_+,d_-,Y_I,m_b^{(0)},m_t^{(0)})$, where $m_b^{(0)}$ and
$m_t^{(0)}$ are the
tree level quark masses for the bottom and top quarks respectively. Expressions for
the eigenvectors and eigenvalues of the mass matrices involved in the numerical evaluation were
obtained using MATHEMATICA, and the $\chi^2$ function was minimized using
MINUIT from the CERNLIB packages\cite{minuit}; both Monte Carlo and standard
routines
were used in the minimization process. The tree level masses of the top
($m_t^{(0)}$) and bottom ($m_b^{(0)}$) quarks were restricted to be around the
central values $\pm$ 10 \%  in order to assure consistency with the assumption
that radiative corrections are small.
The $\chi^2$ function presents an even symmetry with respect to 5 of the
parameters of the matrices in Eq. (\ref{emm}); we find that there are
32 parameter domains
where the $\chi^2$ function takes small values. For the extremal points
this even symmetry is not an exact one,  but all the zones have a  
Yukawa constant of the order of 10 and give masses and CKM matrix elements 
in good agreement with the available experimental values.
The numerical results on one of the 32 minima of the ten parameter space  
are shown in Table \ref{mass:val}.  We use those values (which minimize
$\chi^2$) to calculate, in the context of our model, the fifteen predictions for
$m_q(m_t)$ for
$q=u,d,c,s,t,b$ and $(CKM)_{ij}$ for $i,j=1,2,3$. The numerical results are shown
in Table \ref{ten:res}.

For the sake of comparison, we repeat the same calculations but now using the
central values of the six quark masses at the $M_Z$ scale\cite{fritzsch}. The
numerical results are shown in Table \ref{nein:res}.

\begin{table}
\centering
\begin{tabular}{|c|c|c|c|} \hline 
 &\em RANGE  &\em INPUT &\em BEST FIT \\ \hline
$m_d(m_t)$ & 3.85-5.07 MeV & 4.46 MeV & 3.63 MeV\\
$m_s(m_t)$ & 76.9-100 MeV  & 88.4 MeV & 44.9 MeV \\
$m_b(m_t)$ & 2.74-2.96 GeV &2.85 GeV & 2.91 GeV \\
$m_u(m_t)$ & 1.8-2.63 MeV. & 2.21 MeV& 2.22 MeV \\
$m_c(m_t)$ & 587-700 MeV & 643 MeV&  841 MeV \\
$m_t(m_t)$ & 159-183 GeV & 171 GeV  & 166.5 GeV \\
$CKM_{11}$ & 0.9745-0.9760 &0.9752 & 0.9761  \\
$CKM_{12}$ & 0.217-0.224 &0.2200 &   0.2179   \\  
$CKM_{13}$ & 0.0018-0.0045 &0.0034 &  0.0032 \\  
$CKM_{21}$ & 0.217-0.224 &0.2200 &    0.2214   \\
$CKM_{22}$ & 0.9737-0.9753 & 0.9755 &  0.9742   \\
$CKM_{23}$ & 0.036-0.042 & 0.0390 &    0.0382  \\
$CKM_{31}$ & 0.004-0.013 &0.0085 &    0.0117   \\
$CKM_{32}$ & 0.035-0.042 & 0.0385 &    0.0365   \\
$CKM_{33}$ & 0.9991-0.9994 & 0.9992 &  0.9992  \\
\hline
\end{tabular}
\caption{\label{ten:res}Experimentally allowed values for $m_q(m_t)$ and
CKM matrix
elements. We
show the input and calculated values in the context of our model.}
\end{table}
\hfill
\begin{table}
\centering
\begin{tabular}{|c|c|c|c|} \hline
 &\em RANGE &\em INPUT &\em BEST FIT \\ \hline
$m_d(M_Z)$ & 1.8-5.3 MeV &3.55 MeV  & 3.44 MeV\\
$m_s(M_Z)$ & 35-100  MeV &67.5 MeV  & 39.6 MeV \\
$m_b(M_Z)$ & 2.8-3.0 GeV &2.9 GeV  & 2.9 GeV \\
$m_u(M_Z)$ & 0.9-2.9 MeV &1.19 MeV & 2.03 MeV \\
$m_c(M_Z)$ & 530-680 MeV & 605 MeV & 793 MeV \\
$m_t(M_Z)$ & 168-180 GeV &174 GeV & 166.9 GeV \\
$CKM_{11}$ & 0.9745-0.9760 & 0.9752  & 0.9762    \\
$CKM_{12}$ & 0.217-0.224   & 0.2205  &   0.2174   \\
$CKM_{13}$ & 0.0018-0.0045 & 0.0036 &  0.0030   \\
$CKM_{21}$ & 0.217-0.224 & 0.2205 &    0.2215   \\
$CKM_{22}$ & 0.9737-0.9753 & 0.9745 &  0.9741   \\
$CKM_{23}$ & 0.036-0.042   & 0.0390 &    0.0387   \\
$CKM_{31}$ & 0.004-0.013 & 0.0085 &    0.0116   \\
$CKM_{32}$ & 0.035-0.042 & 0.0385 &    0.0370   \\
$CKM_{33}$ & 0.9991-0.9994 & 0.9992 &  0.9992  \\
\hline
\end{tabular}
\caption{\label{nein:res}Experimentally allowed values for $m_q(M_Z)$ and
CKM matrix elements. We
show the input and calculated values in the context of our model.} 
\end{table}

Let us make two comments: First, the values for the parameters in the scalar field
square mass matrices are of order $10^{17} (\mbox{MeV})^2$ (see Table
\ref{mass:val}), so, the scalar physical masses are of order $10^{3}$ TeV.
Second, the rounding errors allow us to take safely up to five significative
figures in the masses and in the CKM matrix elements

\vskip.3cm

As can be seen from Tables \ref{ten:res} and \ref{nein:res}, even under the
assumption that the CKM matrix elements are real, the numerical values are in
good agreement with the allowed experimental results.

\begin{table}[!ht]
\centering\begin{tabular}{|c|c|c|c|c|c|}\hline
\em PARAMETER &\em $a_{-}$ &\em b &\em $c_{-}$&\em $d_{-}$& \em $c_+$  \\\hline
VALUE& 22.845\f & 1.850 \f & 10.018 \f & -10.227 \f & 13.571 \f  \\\hline
\em PARAMETER & \em $d_{+}$& $a_{+}$ & \em Y & \em $m^{(0)}_b$ & \em $m^{(0)}_t$
\\\hline
VALUE&  13.762 \f& 150.01\f& 13.6 & 2.912 GeV & 166.1 GeV. \\\hline
\end{tabular}
\caption{\label{mass:val}Values of the parameters in one of the minima  
(the values of the scalar mass matrices elements  are in $(\mbox{MeV})^2$
units).}
\end{table}

\section{CONCLUSIONS}
By introducing a $U(1)_X$ gauge flavor symmetry and enlarging the scalar
sector, we have presented a mechanism and an explicit model able to generate
radiatively the hierarchical spectrum of quarks
masses and CKM mixing angles. The horizontal charge assignment to
particles is such that we do not need to go beyond the known three
generations of quarks and leptons. Also, at tree level only the t and b
quarks get masses. To generate radiatively the masses for the light families we
have introduced some new exotic scalars. All of these new scalars are
charged and color non-singlets, so they can not get VEV as is required in
the loop graphs.

\vskip.3cm

Our numerical results are presented in Tables \ref{ten:res} and \ref{nein:res}.
Even though we are guessing the $U(1)_X$ mass scale, the two sets of results do
not differ by much and they agree fairly well with the experimental
values, meaning that the mass scale associated with the horizontal symmetry may
be in the range $100 GeV < M_X< 1.0 TeV$. A closer look to our analysis shows
that we are translating the quark mass hierarchy to
the quotient  $v_1/v_2$ which is the hierarchy between the electroweak mass scale
and the Horizontal $U(1)_X$ mass scale. In this way we demonstrate the viability
that new physics at the electroweak mass scale, or just above it, may help to
explain the long-lasting puzzle of the enormous range of quark masses and
mixing angles.

\vskip.3cm

Since quarks carry baryon number $B=1/3$, the color sextet scalars we
have introduced must have $B=-2/3$ (the scalar singlets $\phi_1$ and
$\phi_2$ have
B=0); in this way $L_{Y_M}$ is not only $U(1)_X$ invariant but conserves color
and baryon number as well. On the other hand, $V(\phi)$ does not
conserve baryon number; as a matter of fact, the  term
$\lambda_4\phi_5\phi_6\phi_7\phi_2$  violates
baryon number by two
units ($\Delta B=\pm 2$) and could induce neutron-antineutron oscillations. A
roughly
estimate of such oscillations shows that they are proportional to
$v_2(M^2_{\phi_5}M^2_{\phi_6}M^2_{\phi_7})^{-1}\sim 10^{-19} GeV$ which is
negligible in principle. Any way, in the worse of the situations, since 
 the  offending term  does not enter on the mass matrix for the Higgs scalars
(it is just there), it may be removed in
more realistic models by the introduction of a discrete symmetry. 

\vskip.3cm

Our results are encouraging; even under the assumption that the CKM matrix is real,
and without knowing exactly the $U(1)_X$ mass scale, the numerical
predictions are in the ballpark, implying also a value of order 10 for 
the Yukawa coupling $Y_I$, and masses for the exotic
scalars being of order $10^3$ TeV. Our model presents thus a clear
mechanism able to explain the mass hierarchy and mixing of the quarks.

\vskip.3cm

Finally let us mention that in the work presented here, the Higgs scalar used to produce the SSB
of the SM gauge
group down to $SU(3)_C\times U(1)_Q$ has zero horizontal charge, and as a
consequence the standard $Z$ boson does not mix with the horizontal
counterpart. However, due to recent interest \cite{langacker} on the
phenomenology of a $Z'$, it is worth to study the possibility of
allowing this mixing in future work.

\begin{acknowledgments}
This work was partially supported by Conacyt in Mexico and Colciencias
and BID in Colombia.

One of us (A. Z.) acknowledges the hospitality of the theory group at CERN
and useful conversations with Marcela Carena and Jean Pestieau.
\end{acknowledgments}

\end{spacing}
\end{document}